\documentclass[12pt]{article}
\usepackage{amssymb,amsmath,epsfig}
\allowdisplaybreaks
\begin{document}

\title{\bf Stability of Oscillating Gaseous Masses in Massive Brans-Dicke Gravity}

\author{M. Sharif $^1$ \thanks{msharif.math@pu.edu.pk} and Rubab
Manzoor $^{1,2}$
\thanks{rubab.manzoor@umt.edu.pk}\\
$^1$ Department of Mathematics, University of the Punjab,\\
Quaid-e-Azam Campus, Lahore-54590, Pakistan.
\\$^2$ Department of Mathematics,\\
University of Management and Technology,\\
Johar Town Campus, Lahore-54782, Pakistan.}
\date{}
\maketitle

\begin{abstract}
This paper explores the instability of gaseous masses for the radial
oscillations in post-Newtonian correction of massive Brans-Dicke
gravity. For this purpose, we derive linearized perturbed equation
of motion through Lagrangian radial perturbation which leads to the
condition of marginal stability. We discuss radius of instability of
different polytropic structures in terms of the Schwarzschild
radius. It is concluded that our results provide a wide range of
difference with those in general relativity and Brans-Dicke gravity.
\end{abstract}

\section{Introduction}

The study of evolution and formation of stellar structures has been
issue of great interest in gravitational physics and cosmology. In
this context, the phenomenon of dynamical stability of celestial
objects has important implications in the analysis. It is believed
that different stability ranges for stellar bodies lead to different
phases of evolution or structure formation of the astronomical
models. In general relativity (GR), Chandrasekhar \cite{2,3,4} was
the first who described a mechanism to explain dynamical instability
of stellar structure in weak field approximation (at post-Newtonian
(pN) limits). He used equation of state involving adiabatic index
$(\gamma)$ and concluded that the fluid remains unstable for
$\gamma<\frac{4}{3}$. Herrera et al. \cite{5} investigated dynamical
evolution of self-gravitating fluids in different configurations
(anisotropic fluid, adiabatic, non adiabatic as well as shearing
viscous fluid). Sharif and his collaborators \cite{6} also explored
characteristics of different celestial fluid configurations in weak
regimes through stability analysis.

The mystery of accelerating expansion of the universe has taken a
remarkable attention in the last decade. In this context, the
mechanism of modified theories of gravity has become a fascinated
candidate. Modified theory of gravity means theory of gravity
followed by modified Einstein-Hilbert action. The viability of these
theories is an issue of great importance. For this reason, these
theories are tested on different gravitational scales such as strong
as well as weak field gravitational regime \cite{1}. In this regard,
the evolution and formation of celestial structure are considered to
be the most suitable test-beds for modified theories. It is believed
that modification of GR introduces some new astrophysical insights
which can explain hidden parts of the universe. In this context, a
large number of researchers have discussed modified astrophysical
analysis \cite{6*}. Nutku \cite{7} studied modified fluid
hydrodynamics that affects the results of Chandraskhar. Recently, we
have discussed modified dynamics of self-gravitating system in both
weak and strong fields \cite{8}.

Brans-Dicke (BD) gravity (natural generalization of GR) \cite{9} is
one of the most explored examples of modified theory which is
considered as a solution of many cosmic issues. This theory modifies
the Einstein-Hilbert action according the Dirac hypothesis, i.e., it
allows dynamical gravitational coupling (converts Newtonian
gravitational constant into dynamical one) by means of dynamical
massless scalar field $(G=\frac{1}{\phi})$. In this gravity,
gravitational effects are described by coupling a massless scalar
field $\phi$ with the curvature part (Ricci scalar). One of the main
features of this theory is that it contains a constant tuneable
parameter $\omega_{BD}$ which is a coupling constant and can adjust
required results. This theory provides suitable solutions of various
cosmic problems but remains unable to probe "graceful exist``
problem of old inflationary cosmology. The inflationary phenomenon
described by BD gravity shows unacceptably large microwave
background perturbations (by collisions between big bubbles) which
can be controlled with the help of specific values of coupling
parameter $\omega_{BD}\leq25$ \cite{9a}. But these defined ranges of
parameter are in conflict with observational limits \cite{9b}.

In order to solve this problem, a massive scalar field is introduced
in the framework of BD gravity \cite{9c} via a potential function
$V(\phi)$. This new gravity is known as massive BD (MBD) gravity or
self-interacting BD gravity. Moreover, BD gravity investigates all
strong field issues (cosmological issues) for negative and small
values of $\omega_{BD}$ \cite{10} but satisfies all weak field tests
(related to solar system) for large and positive values of
$\omega_{BD}$ \cite{11}. The MBD gravity provides a consistency with
weak field gravitational test, i.e., explains cosmic acceleration
for positive and large values of $\omega_{BD}$ \cite{12}. There has
been a large body of literature which describes dynamics of MBD
gravity in many cosmic issues \cite{13a, 13b}. Olmo \cite{11a}
calculated pN limits of MBD equations but he converted only
lowest-order (order of $c^{-2})$ limits of solutions in terms of
potential functions to explore $f(R)$ gravity as a special case of
scalar-tensor gravity. Recently, we have explored hydrodynamics of
different celestial configurations in complete pN correction of MBD
gravity that modify the results of GR and BD gravity \cite{14}.

In this paper, we investigate gaseous system in MBD gravity and
compare the results with GR and BD gravity. For this purpose, we
explore stability of gaseous masses for radial oscillations in weak
field approximation of MBD gravity. The paper is organized as
follows. The next section represents complete pN approximation of
MBD theory in terms of potential as well as super-potential
functions and the dynamical equations. Section \textbf{3} explores
instability of gaseous systems for radial oscillations by means of
Lagrangian perturbation and variational principle. In section
\textbf{4}, we evaluate instability conditions of different
polytropes in MBD theory. Finally, section \textbf{5} summarizes the
results.

\section{Massive Brans-Dicke Gravity and Dynamical Equations}

The action of MBD gravity with ($\kappa^{2}=\frac{8\Pi}{c^{2}}$)
\cite{12} is given by
\begin{equation}\label{1}
S=\frac{1}{2\kappa^{2}}\int d^{4}x\sqrt{-g} [\phi
R-\frac{\omega_{BD}}{\phi}\nabla^{\alpha}{\phi}\nabla_{\alpha}{\phi}-V(\phi)]
+L_{m},
\end{equation}
where $L_{m}$ represents matter distribution depending upon metric.
By varying the above action with respect to $g_{\alpha\beta}$ and
$\phi$, we obtain MBD equations as follows
\begin{eqnarray}\nonumber
G_{\alpha\beta}&=&\frac{\kappa^{2}}{\phi}T_{\alpha\beta}+[\phi_{,\alpha;\beta}
-g_{\alpha\beta}\Box\phi]+\frac{\omega_{BD}}{\phi}[\phi_{,\alpha}\phi_{,\beta}
-\frac{1}{2}g_{\alpha\beta}\phi_{,\mu}\phi^{,\mu}]
-\frac{V(\phi)}{2}g_{\alpha\beta},\\\label{2}
\\\label{3}
\Box\phi&=&\frac{\kappa^{2}T}{3+2\omega_{BD}}
+\frac{1}{3+2\omega_{BD}}[\phi\frac{dV(\phi)}{d\phi}-2V(\phi)],
\end{eqnarray}
where $T_{\alpha\beta}$ shows the energy-momentum tensor,
$T=g^{\alpha\beta}T_{\alpha\beta}$ and $\Box$ represents the
d'Alembertian operator. Equations (\ref {2}) and (\ref{3}) indicate
MBD field equations as well as evolution equation for the scalar
field, respectively. We assume matter distribution as a perfect
fluid which can be compatible with pN regime
\begin{equation}\label{4}
T_{\alpha\beta}=[\rho
c^2(1+\frac{\pi}{c^2})+p]u_{\alpha}u_{\beta}-pg_{\alpha\beta},
\end{equation}
where $\rho,~\rho\pi,~p,~u_{\alpha}$ indicate density,
thermodynamics density, pressure and four velocity, respectively.

\subsection{Post-Newtonian Approximation}

The weak-field limits of any relativistic theory explain the order
of deviations of the local system from its isotropic and homogenous
background. The parameterized pN approximations are widely used as
weak field approximated solutions that are obtained by using the
following Taylor expansion of the metric functions \cite{a1}
\begin{eqnarray}\nonumber
g_{\alpha\beta}&\approx&\eta_{\alpha\beta}+h_{\alpha\beta},\nonumber
\end{eqnarray}
with
\begin{equation}\nonumber
h_{00}\approx h^{(2)}_{00}+h^{(4)}_{00},\quad h_{0i}\approx
h^{(3)}_{0i},\quad h_{ij}\approx h^{(2)}_{ij}.
\end{equation}
Here $\eta_{\alpha\beta}$ shows the Minkowski metric (describing
isotropic and homogenous background of $g_{\alpha\beta}$),
$h_{\alpha\beta}$ indicates deviation of $g_{\alpha\beta}$ from
background values $(\eta_{\alpha\beta})$, $i,j=1,2,3$ and the
superscripts $(2),~(3)$ and $(4)$ describe approximation of order
$(c^{-2}),~(c^{-3})$ as well as $(c^{-4})$. In this approximation
scheme, the field equations are solved formally and the metric
functions are expressed as a sequences of pN functions of source
variables (source of metric function like matter) coupled to
coefficients (pN parameter). These coefficients are based upon the
matching conditions between the local system and cosmological models
or on other constants of the theory. The pN functions are basically
metric potentials which are chosen under reasonable assumption of
Poisson's equations and gauge conditions to have unique solutions
according to pN order of correction \cite{a1}.

In order to discuss stability of gaseous system in MBD gravity and
check the compatibility of our results with the analysis of GR
\cite{2,3}, we approximate the system in pN limits. For this
purpose, we use complete pN approximations (upto order of
$(c^{-4})$) of MBD gravity. The parameterized pN limits of MBD
solutions has been evaluated by using the following expansion of
metric and dynamical scalar field \cite{11a,14}
\begin{eqnarray}\nonumber
g_{\alpha\beta}&\approx&\eta_{\alpha\beta}+h_{\alpha\beta},\\\nonumber
\phi&\approx&\phi_{0}(t_{0})+\varphi^{(2)}(t,x)+\varphi^{(4)}(t,x),\\\nonumber
V(\phi)&\approx& V_{0}+\varphi
\frac{dV_{0}}{d\phi_{0}}+\varphi^{2}\frac{d^{2}V_{0}}{d^{2}\phi_{0}}+....
\end{eqnarray}
Here $t_{0}$ represents time of isotropic and homogenous background
of local system. The term $\phi_{0}=\phi(t_{0})$ shows unperturbed
or initial value of scalar field in isotropic and homogenous
background of local system which vary very slowly with respect to
$t_{0}$. This implies that the cosmological considerations would
allow a slow evolution of $\phi_{0}$ on cosmological timescales.
Since these timescales are much larger than the solar system
timescales, so its evolution may be ignored for physical setup in
weak-field and it is considered as constant. The term
$V_{0}=V(\phi_{0})$ shows the potential function of scalar field at
$t_{0}$ and $\varphi(t,x)$ is the local deviation of scalar field
from $\phi_{0}$.

The parameterized pN approximations of MBD solutions are given by
\cite{14}
\begin{eqnarray}\label{c}
g_{ij}&\approx&(-1-\frac{2\gamma_{BD}
U}{c^{2}}-\frac{\Lambda_{BD}r^{2}}{3c^{2}})\delta_{ij},\\\nonumber
g_{00}&\approx&1-\frac{2U}{f(r)c^{2}}+\frac{1}{2c^{4}}\left[(-2U
+\frac{\Lambda_{BD}r^{2}}{3})^{2}-
\left(\frac{-2U-A(r)}{3+2\omega_{BD}+A(r)}\right)^2\right]\\\label{a''}
&-&2(\Phi+\psi),\\\label{b''}
g_{0i}&\approx&\frac{1}{c^{3}}\left(\frac{4U_{i}}{f(r)}
-\frac{1}{2}\frac{\partial^{2}\chi}{\partial t\partial
x_{i}}\right).
\end{eqnarray}
Here $U=G_{eff}\frac{M_{\odot}}{r}$ ($M_{\odot}=\int
d^{3}x^\ast\rho_{sun}(t,x^\ast)$ is the Newtonian mass of the sun)
is the effective gravitational potential determined by Poisson's
equation
\begin{equation}\label{d}
\nabla^{2}U=-4\Pi\rho G_{eff},
\end{equation}
where $G_{eff}$ indicates the effective gravitational constant
(dynamical Newtonian gravitational constant) for massive scalar
field defined by \cite{13b,11a}
\begin{equation}\label{d'}
G_{eff}=\frac{\kappa^{2}}{8\Pi\phi_{0}}f(r)=\frac{\kappa^{2}}{8\Pi\phi_{0}}\left(1
+\frac{A(r)}{3+2\omega_{BD}}\right),
\end{equation}
where
$$A(r)=\left\{\begin{array}{ll}
e^{-m_{0}r}&\quad m_{0}^{2}> 0 \\
\cos(m_{0}r)&\quad m_{0}^{2}< 0,
\end{array}\right.\quad~
m_{0}=\left(\frac{\phi_{0}\frac{d^{2}V_{0}}{d\phi^2_0}
-\frac{dV_{0}}{d\phi_0}}{3+2\omega_{BD}}\right)^{1/2}.$$ Here the
term $m_{0}$ is the mass of the massive scalar field and ``$r$''
represents scale of experiments and observations. It is actually a
distance between two points in the local system and can be used to
express radius of configuration (spherical or cylindrical) under
consideration. The term $\gamma_{BD}$ represents the parameterized
pN parameter given by
\begin{equation}\label{0}
\gamma_{BD}=\frac{3+2\omega_{BD}-A(r)}{3+2\omega_{BD}+A(r)}.
\end{equation}
The oscillatory solutions $A(r)=\cos(m_{0}r),~ m_{0}^{2}< 0$ are
unacceptable \cite{F}. In this case, the inverse-square law modifies
as
\begin{equation}\label{01}
\frac{M_{\odot}}{r^{2}}\rightarrow\left(1+\frac{\cos(m_{0}r)+(m_{0}r)\sin(m_{0}r)}{}\right)\frac{M_{\odot}}{r^{2}},
\end{equation}
and for very light fields (showing long-range interactions), the
arguments of cosine and sine are very small in solar system scales
$(m_{0}r<<1)$ which provide $\cos(m_{0}r)\approx1$ and
$\sin(m_{0}r)\approx0$. These approximations lead to usual Newtonian
limits upto an irrelevant redefinition of Newtonian Constant. This
also yields $\gamma_{BD}\approx1/2$ for $\omega_{BD}=0$ which is
observationally unacceptable since $\gamma_{obs}\approx1$. If the
scalar interaction is short-range or mid-range, the Newtonian limits
would dramatically be modified. In fact, the leading order term is
then oscillating, $\sin(m_{0}r)\frac{M_{\odot}}{r}$, and is clearly
incompatible with observations. That is why, we consider only the
damped solutions $A(r)=e^{-m_{0}r},~ m_{0}^{2}> 0$.

The Yukawa-type correction in the Newtonian potential has not been
observed over distances that range from meters to planetary scales.
In addition, since the post-Newtonian parameter $\gamma_{BD}$ is
observationally very close to unity, the  mass function present in
Eqs.(\ref{0}) and (\ref{01}) satisfy the constraint
$m_{0}>>\frac{1}{\tilde{r}}$ ($\tilde{r}$ shows the scale of the
observations or experiments testing the scalar field). For solar
system scale observations, the relevant scale is the Astronomical
Unit $(\tilde{r}\simeq AU\simeq10^{-8}km)$ corresponding to a mass
scale $m_{AU}\simeq10^{-27} GeV$. Although this scale is small for
particle physics considerations, but it is still much larger than
the Hubble mass scale $m_{H0}\simeq10^{-42} GeV$ required for
nontrivial cosmological evolution of $\phi$ \cite{G}. Current solar
system constraints \cite{a1,F} of the parameter $\omega_{BD}$ have
been obtained under one of the following assumptions \cite{11a,13b}
\begin{itemize}
\item When the background value of $m_{0}$ is very small
$(m_{0}<<\frac{1}{\tilde{r}})$ (negligible mass of the field) and
$m_{0}<< m_{AU}$, MBD system reduces to simple BD gravity (massive
scalar field becomes massless scalar field) having
\begin{equation}\nonumber
G_{eff}=\frac{\kappa^{2}}{8\Pi\phi_{0}}\frac{4+2\omega_{BD}}{3+2\omega_{BD}},\quad
\gamma_{BD}=\frac{1+\omega_{BD}}{2+\omega_{BD}}.
\end{equation}
That is why the BD theory (massless scalar field) is consistent with
solar system constraints of the Cassini mission for
$\omega_{BD}>40000$.
\item For $m=0$ and $m_{0}\simeq m_{AU}\simeq10^{-27} GeV$, the
observational constraints on $\omega_{BD}$ are same as discussed for
the case $m_{0}<< m_{AU}$.
\item For massive scalar field ($m_{0}>> m_{AU}$ and
$m_{0}>>\frac{1}{\tilde{r}}$), the dynamics of the spatial part of
$\phi$ is frozen on the solar system scale through potential
function of scalar field and all values of $\omega_{BD}$ are
observationally acceptable \cite{a}. It can be noticed that further
limit $(\omega_{BD}\rightarrow\infty)$ reduces the value of
$G_{eff}$ to simple Newtonian gravitational constant $G$ and
$\gamma_{BD}=1$ which is consistent with GR.
\item For $m_{0} \gtrsim 200m_{AU}$, all values of $\omega_{BD}>-\frac{3}{2}$
are observationally allowed.
\end{itemize}

The term $\frac{\Lambda_{BD}}{3c^{2}}=\frac{V_{0}}{6\phi_{0}c^{2}}$
indicates the cosmological term (where $\Lambda_{BD}$ is a
cosmological constant) which is based on the potential of the scalar
field. In order to be consistent with observational data (ranging
from the solar system to clusters of stellar structures), the
contribution due to scalar density should be very small and the
following constraint must be satisfied
\begin{equation}\nonumber
\frac{V_{0}L^{2}}{\phi_{0}}<<1.
\end{equation}
Here $L$ shows the length scale equal to or greater than the solar
system. The term $(\Phi+\psi)$ represents super-potential
$\tilde{\Phi}$ given by the following Poisson's equations \cite{14}
\begin{eqnarray}\label{k}
\nabla^{2}\tilde{\Phi}&=&-4\Pi G_{eff}\rho\sigma,\quad
\tilde{\Phi}=\Phi+2\psi,\\\label{l}
\nabla^{2}\psi&=&-\frac{1}{2\phi_{0}}\left[V_{0}(1+h^{(2)}_{[ij]}
-\frac{\varphi^{(2)}}{\phi_{0}})+\varphi^{(2)}\frac{dV_{0}}{d\phi_{0}}\right],
\end{eqnarray}
here
\begin{equation}\label{k'}
\sigma=\frac{1}{f(r)} \left[\pi+2v^{2}+h^{(2)}_{[ij]}
-\frac{\varphi^{(2)}}{\phi_{0}}+\frac{3p}{\rho}\right],\quad
\nabla^{2}\Phi=-4\Pi
G\rho\tilde{\sigma},\quad\tilde{\sigma}=f(r)\sigma.
\end{equation}
Similarly $\chi$ and $U_{i}$ are potential functions satisfying the
following Poisson's equations
\begin{eqnarray}\label{n}
&&\nabla^{2}\chi=h^{(2)}_{00}=\frac{1}{c^{2}}(-2U+\frac{\Lambda_{BD}r^{2}}{3}),\\\label{o}
&&\nabla^{2}\left(\frac{U_{i}}{f(r)}\right)=-4\Pi G_{eff}\frac{\rho
v_{i}}{f(r)},
\end{eqnarray}
where $\nabla^{2}U_{i}=-4\Pi G \rho v_{i}$ \cite{2}. The effect of
$\phi_{0}$ is taken approximately constant and the effects of
$\dot{\phi_{0}}$ as well as $\ddot{\phi_{0}}$ are neglected. The
solutions satisfy the following gauge condition
\begin{equation}\nonumber
h^{\alpha}_{\mu,\alpha}-\frac{1}{2}h^{\alpha}_{\alpha,\mu}
-\frac{1}{c^{2}\phi_{0}}\frac{\partial\varphi}{\partial x_{\mu}}=0.
\end{equation}
All the assumptions and solutions are consistent with BD gravity in
the limits
$(m_{0}<<\frac{1}{\tilde{r}}),~\frac{V_{0}}{\phi_{0}}\rightarrow0$
\cite{7} and the system reduces to GR with $\omega_{BD}\rightarrow
\infty$ \cite{2}.

\subsection{Hydrodynamics}

According to pN approximation of MBD theory, the equation of
continuity and equation of motion (generalized Euler equation of
Newtonian hydrodynamics) are obtained using
\begin{equation}\label{13}
T^{\alpha\beta}_{;\beta}=0.
\end{equation}
From Eqs.(\ref{a''})-(\ref{13}), the equation of continuity is given
by \cite{2,7,14}
\begin{eqnarray}\nonumber
\frac{\partial \tilde{\rho}}{\partial t}+\frac{\partial}{\partial
x_{i}}\left(\tilde{\rho}v_{i}\right)=0,
\end{eqnarray}
where
\begin{equation}\label{14}
\tilde{\rho}=\rho\left(1+\frac{1}{c^{2}}\left(\frac{1}{2}v^{2}-\frac{\Lambda_{BD}r^{2}}{3}
+\frac{9+6\omega_{BD}-e^{-m_{0}r}}{3+2\omega_{BD}+e^{-m_{0}r}}U\right)\right).
\end{equation}
This shows that the mass function indicated by density
$\tilde{\rho}$ remains conserved. The spatial components of
Eq.(\ref{13}) provide the equation of motion is given by \cite{14}
\begin{eqnarray}\nonumber
&&\frac{\partial \eta v_{i}}{\partial t}+\frac{\partial\eta
v_{i}v_{j}}{\partial x_{j}}+\frac{\partial}{\partial
x_{i}}\left[\left(1+2\gamma_{BD}U+\frac{\Lambda_{BD}
r^2}{3}\right)p\right]+\frac{2\rho}{c^{2}}\frac{d}{dt}
\left[\left(2\gamma_{BD}U\right.\right.\\\nonumber
&&\left.\left.+\frac{\Lambda_{BD}r^2}{3}\right)v_{i}\right]
-\frac{4\rho}{c^2}\frac{d}{dt}
\left[\frac{U_{i}}{f(r)}\right]-\frac{\rho}{c^2}\left[f(r)\sigma\frac{\partial
}{\partial x_{i}}\left(\frac{U}{f(r)}\right)+\frac{\partial
\Tilde{\Phi}}{\partial x_{i}}\right]\\\nonumber
&&-\frac{4\rho}{c^{2}}v_{j}\frac{\partial }{\partial
x_{i}}\left(\frac{U_{j}}{f(r)}\right)
-\frac{\rho}{c^2}\frac{\partial }{\partial
x_{i}}\left(\frac{U}{f(r)}\right)
+\frac{\rho}{2c^{2}}\frac{d}{dt}\left(U_{i}-U_{\alpha;i\alpha}\right)\\\label{15}
&&-\frac{\rho}{2c^{2}} W_{i}+\frac{\rho}{2c^{2}}Z_{i(BD)}=0,
\end{eqnarray}
where $\frac{d}{dt}=\frac{\partial}{\partial t}+\textbf{v}.\nabla$
represents material derivative and
\begin{eqnarray}\nonumber
\frac{\partial^{3}\chi}{\partial^{2}t\partial
x_{i}}&=&\frac{d}{dt}\left(U_{i}-U_{\alpha;i\alpha}\right),\\\nonumber
\eta&=&\rho
\left(1+\frac{1}{c^2}(v^{2}+2U-\frac{2\Lambda_{BD}r^2}{3}
+\pi+\frac{p}{\rho})\right),
\end{eqnarray}
the potential functions $U_{\alpha;i\alpha},~ W_{i}$ and $Z_{i
(BD)}$ are given in Appendix \textbf{A}.

\section{Dynamical Stability of Gaseous Masses}

To discuss stability of gaseous masses in the presence of massive
scalar field, we use Chandrasekhar technique \cite{3} which has been
also used to explain stability of gaseous system in BD gravity
\cite{7}. For this, we assume that initially the spherically
symmetric distribution of matter field is in complete hydrostatic
equilibrium. Using Eq.(\ref{15}), the hydrostatic condition is given
by
\begin{eqnarray}\nonumber
&&\left[\left(1+2\gamma_{BD}\frac{U}{c^{2}}+\frac{\Lambda_{BD}
r^2}{3c^{2}}\right)\right]\frac{\partial p}{\partial x_{i}}=
\frac{\rho}{c^2}\left[f(r)\sigma\frac{\partial }{\partial
x_{i}}\left(\frac{U}{f(r)}\right)+\frac{\partial
\Tilde{\Phi}}{\partial x_{i}}\right]-\frac{\rho}{c^2}\\\label{17}
&&\times\frac{\partial }{\partial
x_{i}}\left(\frac{U}{f(r)}\right)-p\frac{\partial}{\partial
x_{i}}\left[\left(1+2\gamma_{BD}\frac{U}{c^{2}}+\frac{\Lambda_{BD}
r^2}{3c^{2}}\right)\right]-\frac{\rho}{2c^{2}}Z^{h}_{i(BD)},
\end{eqnarray}
where $Z^{h}_{i(BD)}$ represents hydrostatic case of $Z_{i(BD)}$ and
its value is mentioned in Appendix \textbf{A}. In hydrostatic
equilibrium, the values of $\sigma$ and density term $\tilde{\rho}$
are free from velocity term ($v^{2}$).

\subsection{Lagrangian Perturbation and Oscillations}

In order to discuss stability of oscillating MBD fluid, we consider
that the fluid is flowing according to Lagrangian description. In
Lagrangian description of fluid flow, the spatial reference system
is comoving with the fluid. The position of the particle (depending
upon spatial coordinates) is not an independent variable and the
material derivative reduces to simple partial derivative of time at
specific constant position \cite{14'}. We assume that the system is
initially in hydrostatic configuration. Then after certain time, the
equilibrium configuration of the system is slightly perturbed such
that the spherically symmetric distribution remains unchanged. The
perturbed state is obtained by the following Lagrangian displacement
\cite{3,7}
\begin{equation}\nonumber
\bar{\xi} e^{i\alpha t},
\end{equation}
where $\bar{\xi}$ is a displacement vector defined by
$\bar{\xi}=\tilde{\textbf{x}}-\textbf{x},~
(\tilde{\textbf{x}},~\textbf{x}$ representing position vector of
Lagrangian particles from their initial position at time $t$). The
term $\alpha$ shows the characteristic frequency of oscillations. In
order to determine frequency of the oscillations, we evaluate
linearized Lagrangian form of Eq.(\ref{15}) (which governs small
oscillations about the equilibrium) by using lagrangian perturbation
and Eq.(\ref{17}) as follows
\begin{eqnarray}\nonumber
&&\alpha^{2}\left\{\eta\xi_{i}+\frac{2\rho}{c^{2}}
\left(\left(2\gamma_{BD}\frac{U}{c^{2}}+\frac{\Lambda_{BD}}{3c^{2}}r^{2}\right)\xi_{i}
-2\frac{U_{i}}{f(r)}\right)\right.\\\nonumber
&&\left.+\frac{\rho}{c^{2}}(U_{i}-U_{\alpha;i
\alpha})\right\}=-\frac{\partial}{\partial
x_{i}}\left[(1+2\gamma_{BD}\frac{U}{c^{2}}+\frac{\Lambda_{BD}r^{2}}{3c^{2}})\Delta
p+2p\gamma_{BD}\Delta U\right]\\\nonumber &&+\frac{\rho}{c^{2}}
\left[f(r)\sigma\frac{\partial}{\partial x_{i}}\left(\frac{\Delta U
}{f(r)}\right)-f(r)\Delta \sigma\frac{\partial}{\partial
x_{i}}\left(\frac{U}{f(r)}\right)\right]\\\nonumber
&&-\frac{\partial}{\partial x_{i}}\Delta
\Tilde{\Phi}-\frac{\Delta\rho}{\rho}\frac{\partial}{\partial
x_{i}}\left[(1+2\gamma_{BD}\frac{U}{c^{2}}+\frac{\Lambda_{BD}r^{2}}{3c^{2}})p\right]\\\label{18}
&&+\frac{\rho}{c^{2}}\frac{\partial}{\partial
x_{i}}\left(\frac{\Delta U}{f(r)}
\right)+\frac{\rho}{c^2}\xi_{i}\Delta Z_{i(BD)}.
\end{eqnarray}
Here $v_{i}$ is converted into $i\alpha e^{i\alpha t}\xi_{i}$ and
$\frac{d}{dt}$ becomes $\frac{\partial}{\partial t}$. The terms
$\Delta \rho,~\Delta p,~\Delta\sigma,\\~\Delta U,~\Delta Z_{i(BD)}$
and $\Delta \Tilde{\Phi}$ denote the Lagrangian changes in the
respective quantities. The value of $v_{i}$ is replaced by $\xi_{i}$
in the definition of $U_{i}$ as well as in $U_{\alpha;i \alpha}$ and
$\Delta Z_{i(BD)}$ is expressed in appendix \textbf{A}.

Now we express the Lagrangian changes of various dynamical
quantities in terms of $\bar{\xi}$. Under Lagrangian perturbation,
the equation of continuity becomes
\begin{equation}\label{19}
\Delta \tilde{\rho}=-\tilde{\rho}\nabla.\bar{\xi}.
\end{equation}
Equations (\ref{14}) and (\ref{19}) imply
\begin{eqnarray}\nonumber
\Delta \tilde{\rho}&=&\Delta
\rho\left(1+\frac{1}{c^{2}}\left(\frac{\Lambda_{BD}r^{2}}{3}
+\frac{9+6\omega_{BD}-e^{-m_{0}r}}{3+2\omega_{BD}+e^{-m_{0}r}}U\right)\right)
+\frac{\rho}{c^{2}}\\\nonumber
&\times&\frac{9+6\omega_{BD}-e^{-m_{0}r}}{3+2\omega_{BD}+e^{-m_{0}r}}\Delta
U =-\rho\left(1+\frac{1}{c^{2}}\left(-\frac{\Lambda_{BD}r^{2}}{3}
\right.\right.\\\nonumber
&+&\left.\left.\frac{9+6\omega_{BD}-e^{-m_{0}r}}{3+2\omega_{BD}+e^{-m_{0}r}}U\right)\right)\nabla.\bar{\xi}.
\end{eqnarray}
From the above equation, the explicit expressions of Lagrangian
change in density can be evaluated in terms of $\bar{\xi}$ as
\cite{3,7}
\begin{equation}\label{20}
\Delta \rho=-\rho\left(\nabla.\bar{\xi}
+\frac{9+6\omega_{BD}-e^{-m_{0}r}}
{3+2\omega_{BD}+e^{-m_{0}r}}\Delta U\right),
\end{equation}
where only linear terms of $U$ and $\Delta U$ are considered. The
definition of adiabatic index $(\gamma)$ and the relation
\begin{equation}\label{21a}
d\pi=\frac{p}{\rho^{2}}d\rho,
\end{equation}
yield
\begin{equation}\label{22a}
\Delta
p=\gamma\frac{p}{\rho}\Delta\rho,\quad\rho\Delta\pi=\frac{p}{\rho}\Delta\rho.
\end{equation}
Equations (\ref{20}) and (\ref{22a}) give
\begin{eqnarray}\label{21}
\Delta p&=&-\gamma
p\left(\nabla.\bar{\xi}+\frac{9+6\omega_{BD}-e^{-m_{0}r}}
{3+2\omega_{BD}+e^{-m_{0}r}}\Delta U\right),\\\label{22}
\rho\Delta\pi&=&-p \left(\nabla.\bar{\xi}
+\frac{9+6\omega_{BD}-e^{-m_{0}r}}
{3+2\omega_{BD}+e^{-m_{0}r}}\Delta U\right).
\end{eqnarray}
From Eqs.(\ref{k'}) and (\ref{20})-(\ref{22}), it follows that
\begin{eqnarray}\nonumber
\Delta\sigma&=&\frac{1}{f(r)}\left(
\frac{e^{-m_{0}r}}{3+2\omega_{BD}+e^{-m_{0}r}}\Delta
U\right.\\\label{23}
&-&\left.\frac{p}{\rho}(3\gamma-2)\left(\nabla.\bar{\xi}
+\frac{1}{c^{2}}\frac{9+6\omega_{BD}-e^{-m_{0}r}}
{3+2\omega_{BD}+e^{-m_{0}r}}\Delta U\right)\right),
\end{eqnarray}
where only linear terms of $\xi_{i}$ are considered. In order to
obtain explicit expressions of $\Delta U$ and $\Delta \Tilde{\Phi}$
in terms of $\bar{\xi}$, we use relation between Eulerian and
Lagrangian changes given by \cite{3,7}
\begin{eqnarray}\label{23'}
\Delta U=\delta U+\bar{\xi}.\nabla U,\quad \Delta
\Tilde{\Phi}=\delta\Tilde{\Phi}+\bar{\xi}.\nabla\Tilde{\Phi}.
\end{eqnarray}
Here $\delta U$ and $\delta\Tilde{\Phi}$ represent Eulerian changes
in the respective quantities that can be calculated from
Eqs.(\ref{d}) and (\ref{k}) as follows
\begin{equation}\nonumber
\nabla^{2}\delta U=-4\Pi
G_{eff}\delta\rho,\quad\nabla^{2}\delta\Tilde{\Phi}=-4\Pi
G_{eff}\delta(\rho\sigma).
\end{equation}
Integration of the above equation gives \cite{3,7}
\begin{eqnarray}\nonumber
\delta
U&=&\int_{v}G_{(eff)}\rho(\tilde{\textbf{x}})\xi_{i}(\tilde{\textbf{x}})\frac{\partial}{\partial
x_{i}}\frac{1}{|\textbf{x}-\tilde{\textbf{x}}|}d\tilde{\textbf{x}}\\\label{24}
&-&\frac{6}{c^{2}}\int_{v}\frac{3+2\omega_{BD}}{3+2\omega_{BD}+e^{-m_{0}r}}
G_{(eff)}\frac{\rho(\tilde{\textbf{x}})\Delta
U(\tilde{\textbf{x}})}{\textbf{x}-\tilde{\textbf{x}}}d\tilde{\textbf{x}},\\\nonumber
\delta
\Tilde{\Phi}&=&\int_{v}G_{(eff)}\rho(\tilde{\textbf{x}})\sigma(\tilde{\textbf{x}})
\xi_{i}(\tilde{\textbf{x}})\frac{\partial}{\partial
x_{i}}\frac{1}{|\textbf{x}-\tilde{\textbf{x}}|}d\tilde{\textbf{x}}\\\label{25}
&-&\int_{v}G{(eff)}\frac{\rho(\tilde{\textbf{x}})\Delta
\Tilde{\Phi}(\tilde{\textbf{x}})}{\textbf{x}-\tilde{\textbf{x}}}d\tilde{\textbf{x}}.
\end{eqnarray}
Equations (\ref{23'})-(\ref{25}) provide
\begin{eqnarray}\nonumber
\Delta U&=&\bar{\xi}.\nabla U+
\int_{v}G_{(eff)}\rho(\tilde{\textbf{x}})\xi_{i}(\tilde{\textbf{x}})\frac{\partial}{\partial
x_{i}}\frac{1}{|\textbf{x}-\tilde{\textbf{x}}|}d\tilde{\textbf{x}}\\\label{24''}
&-&\frac{6}{c^{2}}\int_{v}\frac{3+2\omega_{BD}}{3+2\omega_{BD}+e^{-m_{0}r}}
G_{(eff)}\frac{\rho(\tilde{\textbf{x}})\Delta
U(\tilde{\textbf{x}})}{\textbf{x}-\tilde{\textbf{x}}}d\tilde{\textbf{x}},\\\nonumber
\Delta \tilde{\Phi}&=&\bar{\xi}.\nabla\tilde{\Phi}+
\int_{v}G_{(eff)}\rho(\tilde{\textbf{x}})\sigma(\tilde{\textbf{x}})
\xi_{i}(\tilde{\textbf{x}})\frac{\partial}{\partial
x_{i}}\frac{1}{|\textbf{x}-\tilde{\textbf{x}}|}d\tilde{\textbf{x}}\\\label{25''}
&-&\int_{v}G{(eff)}\frac{\rho(\tilde{\textbf{x}})\Delta
\tilde{\Phi}(\tilde{\textbf{x}})}{\textbf{x}-\tilde{\textbf{x}}}d\tilde{\textbf{x}}.
\end{eqnarray}
With the help of Eqs.(\ref{20})-(\ref{25''}), Eq.(\ref{18}) can be
expressed explicitly in terms of $\bar{\xi}$.

\subsection{The Variational Principle}

The stability criteria of oscillating body depends upon the behavior
of frequency. For $\alpha^2=0$, the system becomes marginally
stable, i.e., the model will expand and contract with homologous
property. Therefore, in order to discuss the behavior of frequency,
we use variational principle with the help of Eq.(\ref{18}). For
this purpose, we assume that on the boundary $(r=R),~\Delta p=0$ and
at the origin $r=0$, each quantity is nonsingular \cite{3,4}. In
this way, Eq.(\ref{18}) along with boundary conditions represent a
self-adjoint characteristic value problem for $\alpha^2$. Thus a
variational base is obtained by converting Eq.(\ref{18}) into
$\bar{\xi}$ and then integrating over the configuration of fluid by
contracting with $\xi_{i}$ \cite{3}. The resulting equation becomes
\begin{eqnarray}\nonumber
Q\alpha^2&=&\int_{v}(\nabla.\bar{\xi})
p\left[\left(1+2\gamma_{BD}\frac{U}{c^{2}}+\frac{\Lambda_{BD}r^2}{3c^{2}}\right)
\left(\frac{9+6\omega_{BD}-e^{-m_{0}r}}{3+2\omega_{BD}+e^{-m_{0}r}}\Delta
U\right.\right.\\\nonumber&+&\left.\left.\gamma
\nabla.\bar{\xi}\right)+2p\gamma_{BD}\Delta
U\right]d\textbf{x}+\frac{\rho}{c^{2}}\int_{v} \left[f(r)\sigma
\xi_{i}\frac{\partial}{\partial x_{i}}\left(\frac{\Delta U}{f(r)}
\right)-f(r)\Delta \sigma\right.\\\nonumber
&\times&\left.\xi_{i}\frac{\partial}{\partial
x_{i}}\left(\frac{U}{f(r)}
\right)+\xi_{i}\frac{\partial\Tilde{\Phi}}{\partial
x_{i}}\right]d\textbf{x}+\frac{\rho}{c^{2}}\int_{v}\left(\nabla.\bar{\xi}
+\frac{9+6\omega_{BD}-e^{-m_{0}r}}
{3+2\omega_{BD}+e^{-m_{0}r}}\Delta U\right)\\\nonumber &\times&
f(r)\xi_{i}\frac{\partial}{\partial
x_{i}}\left((1+2\gamma_{BD}U+\frac{\Lambda_{BD}r^{2}}{3})p\right)d\textbf{x}
+\frac{\rho}{c^{2}}\int_{v}\xi_{i}\frac{\partial}{\partial
x_{i}}\\\nonumber &\times&\left.\left(\frac{\Delta U}{f(r)}
\right)d\textbf{x}-\frac{2\rho}{c^2}\int_{v}\xi_{i}\frac{\partial}{\partial
x_{i}}\left(\frac{e^{-m_{0}r}}{3+2\omega_{BD}+e^{-m_{0}r}}\Delta U
\right)d\textbf{x}\right.\\\nonumber &+& \left.
\frac{2}{c^2}\int_{v}\left(\gamma
p\left(\nabla.\bar{\xi}+\frac{9+6\omega_{BD}-e^{-m_{0}r}}
{3+2\omega_{BD}+e^{-m_{0}r}}\Delta
U\right)\right.\right.\\\nonumber&\times&\left.\left.\left(\xi_{i}\frac{\partial}{\partial
x_{i}}(\frac{3+2\omega_{BD}+2e^{-m_{0}r}}{3+2\omega_{BD}
+e^{-m_{0}r}} +\gamma_{BD})U\right)\right)
d\textbf{x}\right.\\\nonumber&+&\left.\frac{2p}{c^2}\int_{v}\left(\xi_{i}\frac{\partial}{\partial
x_{i}}(\frac{3+2\omega_{BD}+2e^{-m_{0}r}}{3+2\omega_{BD}+e^{-m_{0}r}})\Delta
U\right)d\textbf{x}\right.\\\label{26}
&+&\left.2\int_{v}\left[\xi_{i}\frac{\partial}{\partial
x_{i}}(\gamma_{BD}\Delta U)+\Delta
U\Lambda_{BD}\xi_{i}\frac{\partial}{\partial
x_{i}}\frac{r^2}{3}\right]d\textbf{x}\right].
\end{eqnarray}
The left hand of this equation is
\begin{eqnarray}\nonumber
Q\alpha^2&=&\alpha^2\left\{\int_{v}\eta|\bar{\xi}|^{2}d\textbf{x}
+\frac{\Lambda_{BD}r^2}{3}|\bar{\xi}|^{2}d\textbf{x}+\int_{v}\int_{v}
\frac{G_{(eff)}}{c^2}\rho(\textbf{x})\rho(\tilde{\textbf{x}})\right.\\\nonumber
&\times&\left.\frac{|\bar{\xi}(\textbf{x})
-\bar{\xi}(\tilde{\textbf{x}})|}{|\textbf{x}-\tilde{\textbf{x}}|}
d\textbf{x}d\tilde{\textbf{x}}\left[2\gamma_{BD}
-\frac{4(3+2\omega_{BD})}{3+2\omega_{BD}+e^{-m_{0}r}}\right]\right.\\\nonumber
&-&\left.\int_{v}\int_{v}\frac{G_{(eff)}}{2}\rho(\textbf{x})
\rho(\tilde{\textbf{x}})
\frac{[\bar{\xi}(\textbf{x}).(\textbf{x}
-\tilde{\textbf{x}})][\bar{\xi}(\tilde{\textbf{x}}).(\textbf{x}-
\tilde{\textbf{x}})]}{|\textbf{x}-\tilde{\textbf{x}}|^3}d\textbf{x}
d\tilde{\textbf{x}}\right\},\\\label{27}
\end{eqnarray}
where $Q$ represents positive-definite quantity.

\subsection{The Onset of Instability for the Radial Oscillations in the
Post-Newtonian Approximation}

Here, we discuss the criteria for the onset of dynamical instability
in pN limits of MBD gravity.  For this purpose, we consider radial
oscillations having density as well as pressure distribution in the
equilibrium conditions. According to definitions of vector spherical
harmonics in radial oscillations, the Lagrangian displacement turns
out to be \cite{3,7}
\begin{eqnarray}\label{28}
\xi_{r}=r\tilde{\eta},\quad \xi_{\perp}=0,\quad \xi_{\theta}=0,
\end{eqnarray}
where ${\tilde{\eta}}$ is an unknown function. The radial components
of $\Delta U,~\Delta \Tilde{\Phi}$ and $\Delta\sigma$ can be
obtained from Eqs.(\ref{23}), (\ref{23'}) and (\ref{28}) as follows
\begin{eqnarray}\nonumber
\Delta\sigma&=&\frac{1}{f(r)}\left[
\frac{e^{-m_{0}r}}{3+2\omega_{BD}+e^{-m_{0}r}}\Delta
U\right.\\\label{32'}
&-&\left.\frac{p}{\rho}(3\gamma-2)\left(\frac{d}{dr}(r^3\tilde{\eta})\right)\right]+
O(c^{-2}),\\\label{29} \Delta U&=&\delta
U+r\tilde{\eta}\frac{dU}{dr},\quad \Delta \Tilde{\Phi}=\delta
\Tilde{\Phi}+r\tilde{\eta}\frac{d\Tilde{\Phi}}{dr}.\\\nonumber
\end{eqnarray}
Here the values of $\delta U$ and $\delta \Tilde{\Phi}$ for radial
oscillations are given by \cite{3}
\begin{eqnarray}\nonumber
\delta U&=&4\Pi G_{(eff)}\left[\int^{R}_{r}\rho(s)s\tilde{\eta}ds
-\frac{3}{c^2}\left(\frac{1}{r}\int^{r}_{0}\rho(s)\Delta U(s)s^{2}
ds\right.\right.\\\label{30}
&+&\left.\left.\int^{R}_{r}\rho(s)\Delta U(s)s
ds\right)\right],\\\nonumber \delta \Tilde{\Phi}&=&4\Pi
G_{(eff)}\left[-\int^{R}_{r}\rho(s)\sigma ds +
\left(\frac{1}{r}\int^{r}_{0}\rho(s)\Delta \sigma(s)s^{2}
ds\right.\right.\\\label{31}
&+&\left.\left.\int^{R}_{r}\rho(s)\Delta \sigma(s)
sds\right)\right].
\end{eqnarray}
Using Eqs.(\ref{17}), (\ref{28})-(\ref{29}) and boundary conditions,
Eq.(\ref{26}) simplifies to
\begin{eqnarray}\nonumber
Q\alpha^2&=&\int^{R}_{0}p\left[1+2\gamma_{BD}\frac{U}{c^{2}}+\frac{\Lambda_{BD}
r^2}{3c^{2}}\right]\left[\gamma
r^{4}(\frac{d\tilde{\eta}}{dr})^{2}+(3\gamma-4)\frac{d}{dr}(r^{3}\tilde{\eta}^{2})\right]dr\\\nonumber
&+&\frac{1}{c^2}\left\{\int^{R}_{0}\rho[\Delta
U]^{2}r^{2}dr+2\int^{R}_{0} (\frac{(9+6\omega_{BD}
-e^{-m_{0}r})\gamma}{3+2\omega_{BD}+e^{-m_{0}r}}-2\gamma_{BD})\right.\\\label{32}
&\times&\left.p\Delta U\frac{d}{dr}(r^{3}\tilde{\eta})dr\right\}.
\end{eqnarray}

The condition for marginal stability will be derived from the above
equation by setting $\alpha^{2}=0$. In particular, for
$\alpha^{2}=0,~\gamma=constant=4/3$ and Newtonian limits (order less
than $c^{-2}$) of equilibrium condition, Eq.(\ref{32}) implies that
$\tilde{\eta}=constant$ (as a solution of the respective equation).
This implies that in Newtonian approximation, the marginal stability
is obtained for $\gamma-4/3=0$ and $\tilde{\eta}=constant$.
Accordingly, in pN approximation, this leads to
\begin{equation}\label{33}
\gamma-4/3=\emph{O}(c^{-2})\quad\tilde{\eta}=constant+\emph{O}(c^{-2}).
\end{equation}
Consequently, from Eq.(\ref{32}) the condition of marginal stability
in pN limits is given by
\begin{eqnarray}\nonumber
&&(3\gamma-4)\int^{R}_{0}p\left(1+2\gamma_{BD}\frac{U}{c^{2}}+\frac{\Lambda_{BD}
r^2}{3c^{2}}\right)\frac{d}{dr}(r^{3}\tilde{\eta}^{2})dr\\\nonumber
&&=-\frac{1}{c^2}\left(\int^{R}_{0}\rho(\Delta
U)^{2}r^{2}dr+\frac{2}{3}\int^{R}_{0}\frac{15+10\omega_{BD}
-e^{-m_{0}r}}{3+2\omega_{BD}+e^{-m_{0}r}}\right.\\\label{34}
&&\times\left.p(\Delta U)r^{2}dr\right),
\end{eqnarray}
where terms upto $\emph{O}(c^{-4})$ are neglected and $\Delta U$ is
approximated as
\begin{equation}\nonumber
\Delta U=-4\Pi G_{(eff)}\int^{R}_{r}\rho s ds + r\frac{dU}{dr}.
\end{equation}
This equation upto $\emph{O}(c^{-2})$ is given by
\begin{eqnarray}\nonumber
9(\gamma-4/3)\int^{R}_{0}pr^{2}dr&=&-\frac{1}{c^2}\left(\int^{R}_{0}\rho(\Delta
U)^{2}r^{2}dr\right.\\\nonumber
&+&\left.\frac{2}{3}\int^{R}_{0}\frac{15+10\omega_{BD}
-e^{-m_{0}r}}{3+2\omega_{BD}+e^{-m_{0}r}}p(\Delta U)r^{2}dr\right).
\end{eqnarray}

From the definition of mass function $M$ and gravitational potential
energy $W$ in equilibrium configuration \cite{3}, the above equation
turns out to be
\begin{eqnarray}\nonumber
(\gamma-4/3)&=&\frac{1}{3c^{2}W}\left(\int^{R}_{0}(\Delta U)^{2}dM
\right.\\\label{35}
&+&\left.\frac{2}{3}\int^{R}_{0}\frac{15+10\omega_{BD}
-e^{-m_{0}r}}{3+2\omega_{BD}+e^{-m_{0}r}} \frac{p}{\rho}\Delta
UdM\right),
\end{eqnarray}
where $dM=4\Pi\rho r^{2}dr$ and $W=-12\Pi\int^{R}_{0}pr^{2}dr$. This
equation describes criteria for the onset of dynamical instability
of gaseous masses in the pN limits of MBD gravity that involve no
information about the equilibrium condition beyond the Newtonian
framework. Notice that this derivation is for a special case
$\gamma=constant$ and the defined criteria depend upon $\omega_{BD}$
as well as mass function $m_{0}$. In the limits
$(m_{0}<<\frac{1}{\tilde{r}})$ and
$\frac{V_{0}}{\phi_{0}}\rightarrow0$, the system reduces to simple
BD, whereas within the limits
$(m_{0}<<\frac{1}{\tilde{r}}),~\frac{V_{0}}{\phi_{0}}\rightarrow0$
and $\omega_{BD}\rightarrow\infty$, the above equation becomes
consistent with GR. The instability criteria obtained in Newtonian
limits are the same as described by theories of GR and BD. However,
in pN limits, the resulting criteria are changed due to the last
term of the above equation.

\section{Dynamical Instability of Polytropes}

Polytropes being self-gravitating spheres represent an approximation
of more relativistic stellar models \cite{15}. In order to obtain
criteria for the onset of dynamical instability of polytropes in
massive gravity, we convert all the quantities $(r,~\rho,~p)$ and
$\Delta U$ into standard Emden variables ($\varepsilon$ and
$\theta$) defined by
\begin{equation}\nonumber
r=\beta\varepsilon,\quad \rho=\rho_{c}\theta^{n},\quad
p=p_{c}\theta^{n+1},
\end{equation}
where $\beta,~\rho_{c},~p_{c}$ represent a scale length, central
density, central pressure and $\theta^{n}=\theta^{n}(\varepsilon)$
is a Lane Emden function with $n$ as a polytropic index.  Under
these conditions, Eq.(\ref{35}) becomes \cite{3,7}
\begin{eqnarray}\nonumber
(\gamma-4/3)&=&-\frac{2G_{(eff)}M}{Rc^{2}}
\frac{(5-n)}{18(n+1)\varepsilon_{1}^{4}|\theta^{'}_{1}|^{3}}
\left\{(n+1)\int^{\varepsilon_{1}}_{0}\theta^{n}\left(\Delta
U(\varepsilon)\right)^{2}\varepsilon^{2}d
\varepsilon\right.\\\label{36}
&+&\left.\frac{2}{3}\int^{\varepsilon_{1}}_{0}\frac{15+10\omega_{BD}
-e^{-m_{0}\beta\varepsilon}}{3+2\omega_{BD}+e^{-m_{0}\beta\varepsilon}}
\theta^{n+1}\Delta
U(\varepsilon)\varepsilon^{2}d\varepsilon\right\}.
\end{eqnarray}
Here $\varepsilon_{1}$ shows the first zero of $\theta^{n}$,
$\theta^{'}_{1}$ represents the value of first derivative of
$\theta^{n}$ at $\varepsilon_{1}$. The above equation describes
conditions for marginal stability of polytropes in MBD gravity,
which depend upon values of $\omega_{BD}$ and $m_{0}$. In order to
analyze some results of physical interest, we apply approximation
scheme (use series solutions of exponential and Lane-Emden function)
on the last term of the above equation and use
\begin{equation}\nonumber
\Delta U(\varepsilon)=-\int^{\varepsilon_{1}}_{\varepsilon}
\theta^{n}\varepsilon
d\varepsilon+\varepsilon\frac{d\theta}{d\varepsilon}
=-(\theta+\varepsilon_{1}|\theta^{'}_{1}|).
\end{equation}
Equation (\ref{36}) provides the resultant conditions for marginal
stability (or criteria of onset of instability) as follows
\begin{equation}\label{37}
R=\frac{K}{\gamma-4/3}R_{s},
\end{equation}
where $R_{s}=\frac{2G_{(eff)}M}{c^{2}}$ is the Schwarzschild radius
and $K$ is the constant term given by
\begin{eqnarray}\nonumber
K&=&\frac{5-n}{18(n+1)\varepsilon^{4}_{1}|\theta^{'}_{1}|}
\left\{\left[2(11-n)\int^{\varepsilon_{1}}_{0}\theta(\frac{d\theta}
{d\varepsilon})^{2}\varepsilon^{2}d\varepsilon+1\right]
-26(1+\omega_{BD})\right.\\\nonumber
&\times&\left.\int_{0}^{\varepsilon_{1}} \theta^{n+2}\varepsilon^{2}
d\varepsilon -26(1+\omega_{BD})\varepsilon^{2}_{1}|\theta^{'}_{1}|
\int^{\varepsilon_{1}}_{0}\theta^{n+1}
\varepsilon^{2}d\varepsilon-(39+26\omega_{BD})
m_{0}\beta\right.\\\label{A}
&\times&\left.\left(\int^{\varepsilon_{1}}_{0}\theta^{n+1}
\varepsilon^{3}d\varepsilon+\int^{\varepsilon_{1}}_{0}\theta^{n+2}
\varepsilon^{3}d\varepsilon\right)\right\}.
\end{eqnarray}
Equation (\ref{37}) shows radius of the system where it becomes
unstable or equivalently the system becomes unstable if the mass of
the system is contracted to radius $R$. If the radius of gaseous
mass is greater than $R$, it remains stable in MBD gravity. Since
the obtained radius of instability is a factor of the Schwarzschild
radius, so the ratio $R/R_{s}$ should be greater than or equal to
zero for real and physical results.

It can be noticed that the  instability analysis depends upon five
parameters. Equation (\ref{37}) describing radius of instability
depends upon the values of $K$ and adiabatic index $\gamma$. The
value of $K$ (given in Eq.(\ref{A})) in turn depends upon the
polytropic index $n$, Lane-Emden function $\beta$, tuneable
parameter $\omega_{BD}$ and mass function $m_{0}$. In order to avoid
complexity, we use fixed values of some parameters. Literature shows
that the instability criteria in GR and BD gravity usually depend
upon adiabatic index $\gamma$ and $\omega_{BD}$, hence we cannot fix
their values. Moreover, in the case of MBD gravity, the behavior of
mass function on the instability criteria is also of great
importance. It has been shown that for $m_{0}\gtrsim 200
m_{AU}=200\times 10^{-27}$, all values of $\omega_{BD}>-\frac{3}{2}$
represent observationally allowed regions \cite{13b}.

Different polytropic indices lead to different stellar structures
out of which configurations defined for $n=0,~1,~1.5,~2,~3$ and
$n<5$ are considered to be realistic stars \cite{15}. Thus, firstly
we evaluate values of radius of instability by calculating $R/R_{s}$
for different polytropic indices ($n=0,~1,~1.5,~2,~3,~5)$ as well as
$n>5$ with fixed value of mass function $m_{0}= 200 m_{AU}=200\times
10^{-27}$ and $\beta=1$. In this way, the resulting instability
criteria will depend upon $\omega_{BD}$ as well as adiabatic index
$\gamma$ and it can be easily comparable with GR \cite{3} and BD
\cite{7} theories. Secondly, we choose a fixed value of
$\omega_{BD}$ (from the extracted instability criteria) along with
$\beta=1$ and find behavior of increasing mass $m_{0}\gtrsim 200
m_{AU}=200\times 10^{-27}$ on instability analysis for polytropes.

\subsection{Polytropes for $n=0$}

Polytropic structures for $n=0$ represent incompressible
configurations in which density remains constant throughout the
surface and pressure varnishes at the surface of stellar structure.
The ranges of instability for this type of star are shown in Figure
\textbf{1}. It can be observed that for
$-1.5\leq\omega_{BD}\leq0.5,~\gamma>4/3$, the obtained radius is
approximately $10$ times more than $R_{s}$. The values
$0.5<\omega_{BD},~0.5<\gamma\leq4/3$ provide valid radii ranges
while $\gamma\geq4/3$ implies un-physical results. In GR and BD
gravity, for $n=0,~\gamma>4/3$, the obtained radii are
$R\approx678.57R_{s}$ and $R\approx660.70R_{s}$, respectively. Thus
for $\gamma>4/3$, the system collapses earlier in MBD gravity than
that in GR and BD gravity.
\begin{figure}\centering \epsfig{file=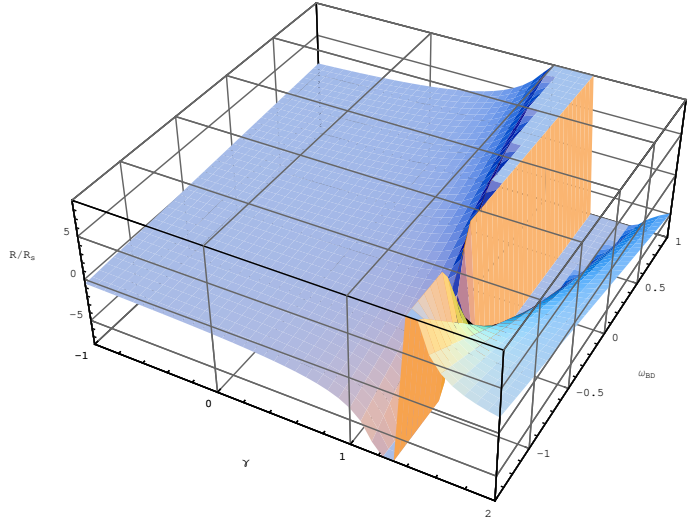,width=.66\linewidth}
\caption{The ratio of radius of instability and Schwarzschild radius
``$R/R_{s}=K/(\gamma-4/3)$" is plotted against $(-1\leq\gamma\leq
2),~ (-1.5\leq\omega_{BD}\leq1)$ for $n=0,~\beta=1$ and $m_{0}= 200
m_{AU}=200\times 10^{-27}$.}
\end{figure}
\begin{figure}\centering
\epsfig{file=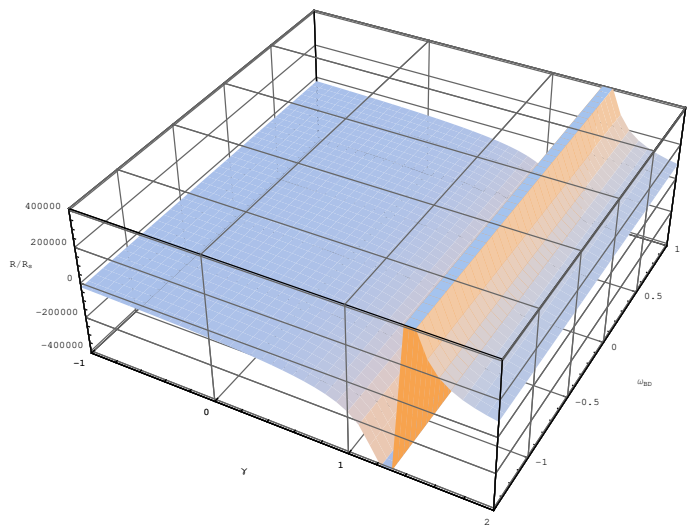,width=.66\linewidth} \caption{The ratio of radius
of instability and Schwarzschild radius ``$R/R_{s}=K/(\gamma-4/3)$"
is plotted against $(-1\leq\gamma\leq 2),~
(-1.5\leq\omega_{BD}\leq1)$ for $n=1,~\beta=1$ and $m_{0}= 200
m_{AU}=200\times 10^{-27}$.}
\end{figure}

\subsection{Polytropes for $n=1$}

Polytropic configurations for $n=1$ show fully convective types of
stars such as neutron stars and very cool late-type stars. The
stability ranges of such type of star are given in Figure
\textbf{2}. It is obvious from the graph that for
$\omega_{BD}>-1.5,~\gamma>4/3,$ the resulting radius of instability
is $R\approx400000R_{s}$ which is much greater than the
Schwarzschild radius. In this case, GR has
$R\approx8.4807\times10^{7}R_{s}$ and BD has $R\approx809.46R_{s}$.
This implies that in MBD gravity, the system becomes unstable before
the time mentioned by GR and much after the time described by BD
gravity.
\begin{figure}\centering
\epsfig{file=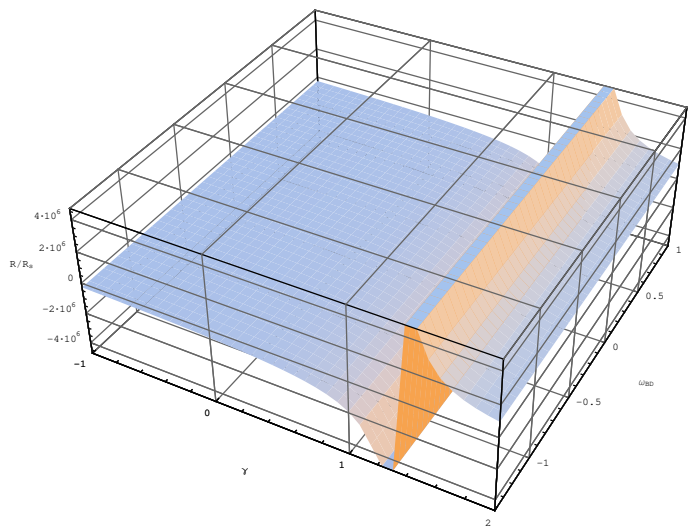,width=.66\linewidth} \caption{The ratio of radius
of instability and Schwarzschild radius ``$R/R_{s}=K/(\gamma-4/3)$"
is plotted against $(-1\leq\gamma\leq 2),~
(-1.5\leq\omega_{BD}\leq1)$ for $n=1.5,~\beta=1$ and $m_{0}= 200
m_{AU}=200\times 10^{-27}$.}
\end{figure}

\subsection{Polytropes for $n=1.5$}

Structures for $n=1.5$ represent good models of stars having fully
convective interior. Figure \textbf{3} represents instability ranges
of such stars under MBD gravity. For $\omega_{BD}>-1.5,~\gamma>4/3$,
the resulting radius is $R\approx1\times10^{6}R_{s}$ while in GR
$R\approx9.67594\times10^{8}$. Thus, for $n=1.5$, the masses become
unstable in MBD much before the limit predicted by GR.
\begin{figure}\centering
\epsfig{file=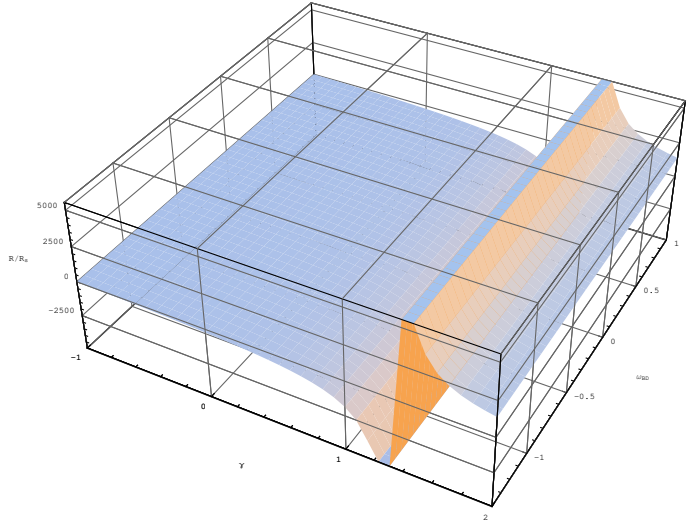,width=.66\linewidth} \caption{The ratio of radius
of instability and Schwarzschild radius ``$R/R_{s}=K/(\gamma-4/3)$"
is plotted against $(-1\leq\gamma\leq 2),~
(-1.5\leq\omega_{BD}\leq1)$ for $n=2,~\beta=1$ and $m_{0}= 200
m_{AU}=200\times 10^{-27}$.}
\end{figure}

\subsection{Polytropes for $n=2$}

The radii of instability for structures $n=2$ are shown in Figure
\textbf{4}. It can be noticed that $\omega_{BD}>-1.5,~\gamma>4/3$
express real results and $R\approx5000R_{s}$. The obtained radii in
GR as well as BD are $R\approx1126.94R_{s}$ and
$R\approx1053.85R_{s}$, respectively. In this case, the radius of
instability in MBD gravity is much greater than that evaluated in GR
and BD theory and hence the system is more stable in MBD gravity.

\subsection{Polytropes for $n=3$}

Polytropic index $n=3$ represents main sequences of stars that have
degenerated cores such as white dwarfs. The stability ranges of this
type of structure in MBD gravity are given in Figure \textbf{5}
which show that for $\omega_{BD}>-1.5,~\gamma>4/3$, the obtained
radius is $R\approx100000R_{s}$. In GR and BD gravity, we have
$R\approx1686.7R_{s}$ and $R\approx1544.1R_{s}$, respectively. Thus
the system is more stable in MBD gravity than in GR and BD theory.
\begin{figure}\centering
\epsfig{file=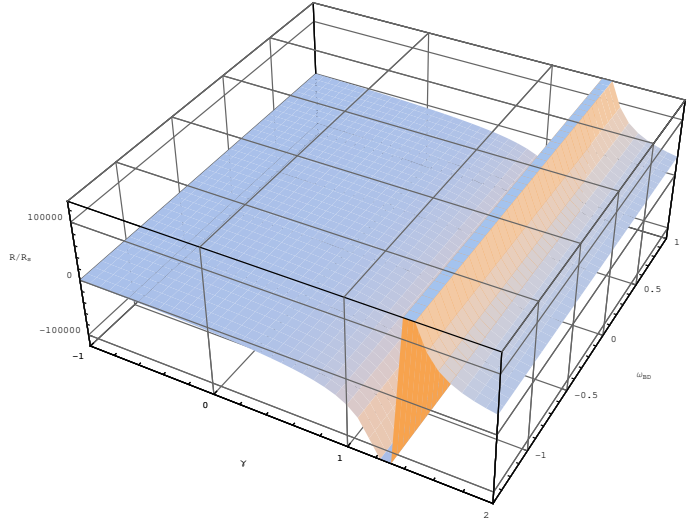,width=.66\linewidth} \caption{The ratio of radius
of instability and Schwarzschild radius ``$R/R_{s}=K/(\gamma-4/3)$"
is plotted against $(-1\leq\gamma\leq 2),~
(-1.5\leq\omega_{BD}\leq1)$ for $n=3,~\beta=1$ and $m_{0}= 200
m_{AU}=200\times 10^{-27}$.}
\end{figure}

\subsection{Polytropes for $n=5$ and $n>5$}

As we have already discussed, various polytropic indices lead to
different stellar configurations out of which structure discussed
for $n=0$ to $n<5$ are proved to be realistic stars \cite{15}.
However, some expected behavior of $n>5$ on instability criteria can
be obtained from Eqs.(\ref{37}) and (\ref{A}). It can be noticed
from Eqs.(\ref{37}) and (\ref{A}) that for $n=5$ we have $K=0$ which
in turn give unphysical result $R=0$. In the case $n>5$, if the
values of $K$ become negative, the radius of instability ($R$)
remains physically acceptable if $\gamma<\frac{4}{3}$, otherwise
$\gamma>\frac{4}{3}$ is the instability criteria.
\begin{figure}\centering \epsfig{file=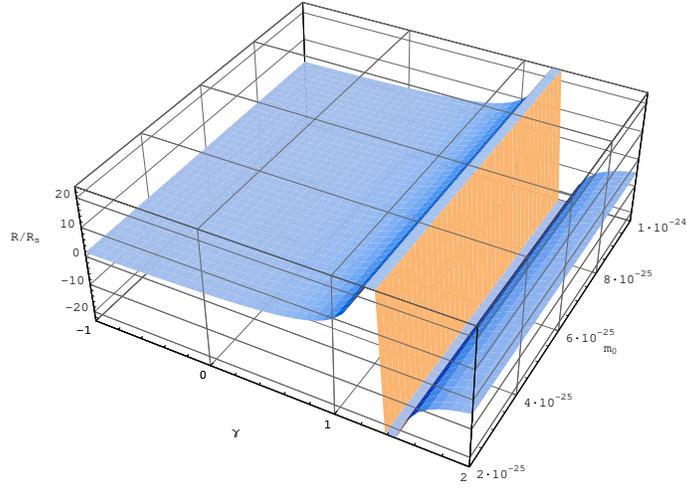,width=.66\linewidth}
\caption{The ratio of radius of instability and Schwarzschild radius
``$R/R_{s}=K/(\gamma-4/3)$" is plotted against $(-1\leq\gamma\leq
2),~(200\times10^{-27}\leq\ m_{0}\leq1000\times10^{-27})$ for
$n=0,~\beta=1$ and $\omega_{BD}=1$.}
\end{figure}
\begin{figure}\centering
\epsfig{file=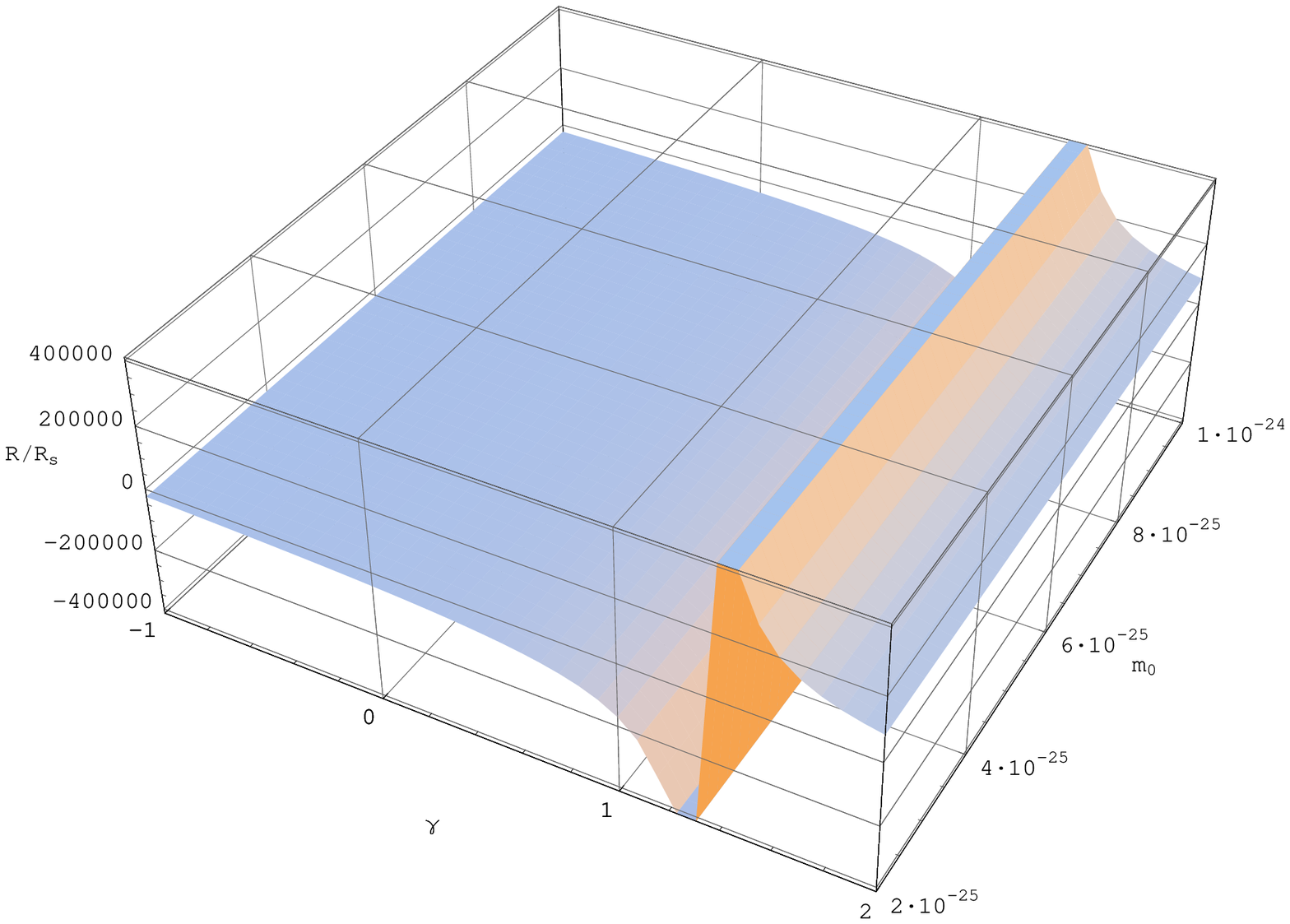,width=.66\linewidth} \caption{The ratio of
radius of instability and Schwarzschild radius
``$R/R_{s}=K/(\gamma-4/3)$" is plotted against $(-1\leq\gamma\leq
2),~(200\times10^{-27}\leq\ m_{0}\leq1000\times10^{-27})$ for
$n=1,~\beta=1$ and $\omega_{BD}=1$.}
\end{figure}
\begin{figure}\centering
\epsfig{file=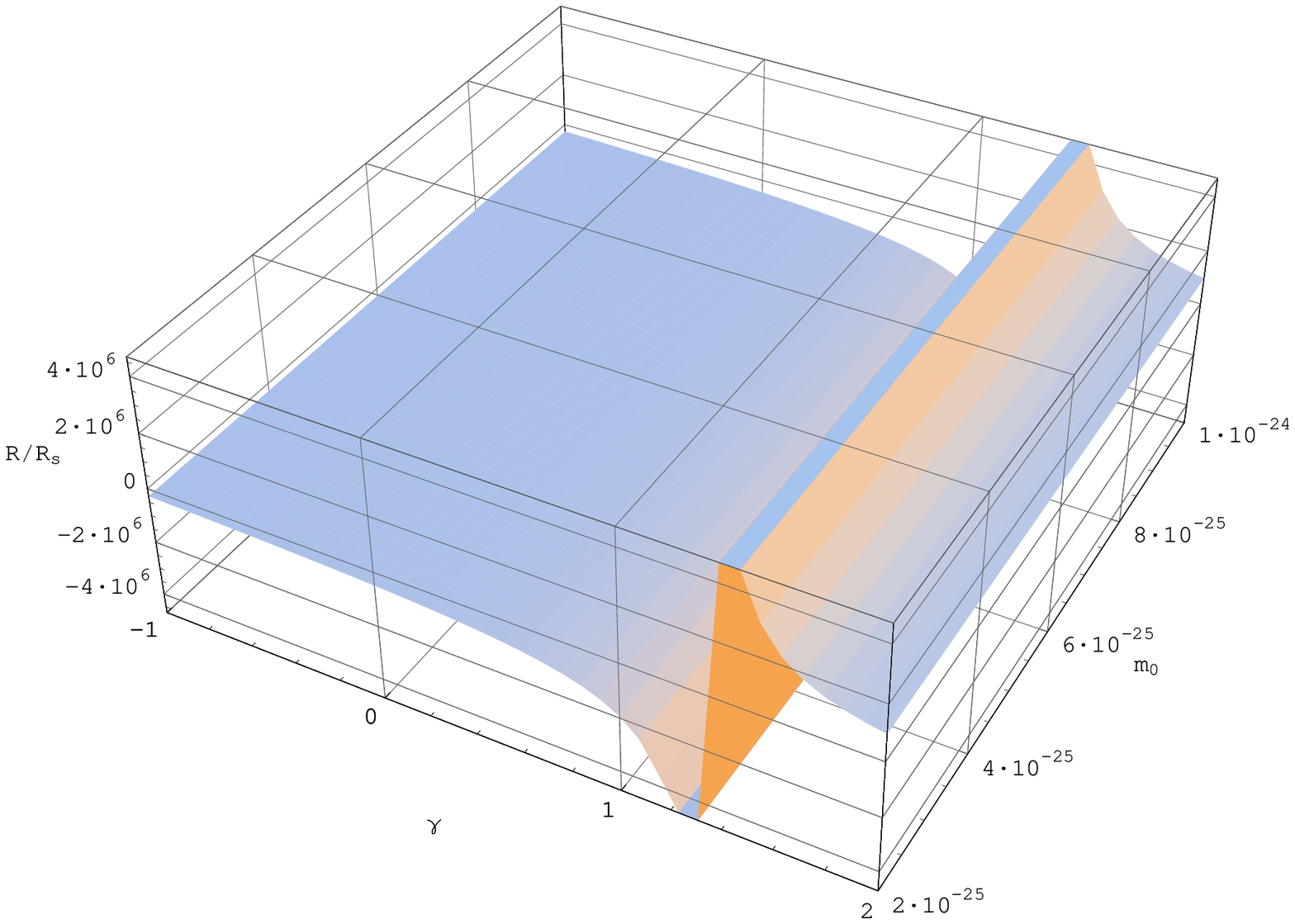,width=.66\linewidth} \caption{The ratio of
radius of instability and Schwarzschild radius
``$R/R_{s}=K/(\gamma-4/3)$" is plotted against $(-1\leq\gamma\leq
2),~(200\times10^{-27}\leq\ m_{0}\leq1000\times10^{-27})$ for
$n=1.5,~\beta=1$ and $\omega_{BD}=1$.}
\end{figure}
\begin{figure}\centering
\epsfig{file=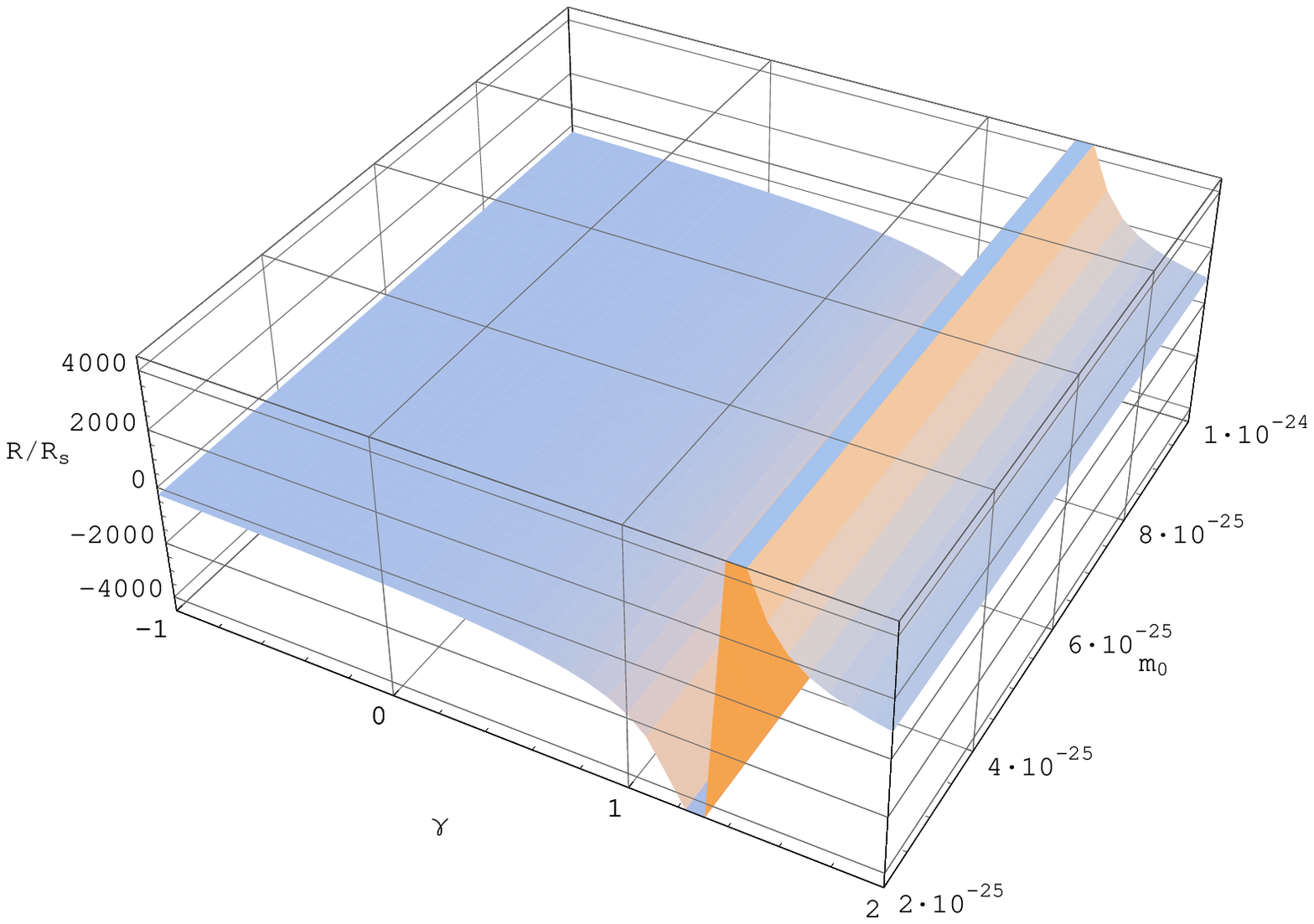,width=.66\linewidth} \caption{The ratio of
radius of instability and Schwarzschild radius
``$R/R_{s}=K/(\gamma-4/3)$" is plotted against $(-1\leq\gamma\leq
2),~(200\times10^{-27}\leq\ m_{0}\leq1000\times10^{-27})$ for
$n=2,~\beta=1$ and $\omega_{BD}=1$.}
\end{figure}
\begin{figure}\centering
\epsfig{file=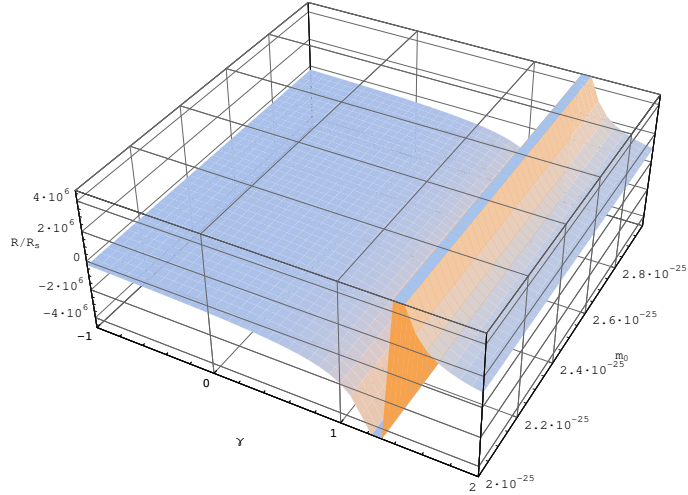,width=.66\linewidth} \caption{The ratio of
radius of instability and Schwarzschild radius
``$R/R_{s}=K/(\gamma-4/3)$" is plotted against $(-1\leq\gamma\leq
2),~(200\times10^{-27}\leq\ m_{0}\leq1000\times10^{-27})$ for
$n=3,~\beta=1$ and $\omega_{BD}=1$.}
\end{figure}

\subsection{Effects of Massive Scalar Field on Stability Criteria}

In MBD gravity, we cannot ignore the behavior of scalar field mass
upon the instability criteria. Figures \textbf{6}-\textbf{11} show
plotting of $R/R_{s}=K/(\gamma-4/3)$ versus increasing mass function
$m_{0}\geq 200 m_{AU}=200\times 10^{-27}$ and $(-1\leq\gamma\leq 2)$
for $n=0,~1,~1.5,~2,~3,$ with $\omega_{BD}=\beta=1$. The value of
$\omega_{BD}$ is chosen from the results of Figure
\textbf{1}-\textbf{5}. It can easily be noticed from these figures
that the variation of mass function does not disturb the behavior of
instability criteria. The constraints on $\gamma$ remains the same
as discussed previously for $n=0,~1,~1.5,~2,~3,$ in Figure
\textbf{1}-\textbf{5}. However, it can be observed that the radii of
instability defined for polytropes in MBD gravity are several orders
of magnitude different from GR and BD theories. This is due to the
coupling of self-interacting massive scalar field with the curvature
term. It is believed that theories of gravity that deviate widely
from GR can lead to the development of suitable modified theory of
gravity. It has been shown that phenomenon in the presence of
massive scalar field (in massive scalar-tensor theories) can differ
drastically from the pure general relativistic one \cite{a}.

The above analysis indicates that all the cases except $n=0,~n=5$
and $n>5$ have stable region for $\gamma>4/3$ which is consistent
with GR and BD gravity. The comparison of instability ranges
($R/R_{s}=K/(\gamma-4/3)$) in MBD, GR \cite{3} and BD \cite{7} are
more clearly described in Figure \textbf{11}. Here, MBD results are
plotted in red, BD ranges are shown in black while purple mapping
describes GR limits.
\begin{figure}\epsfig{file=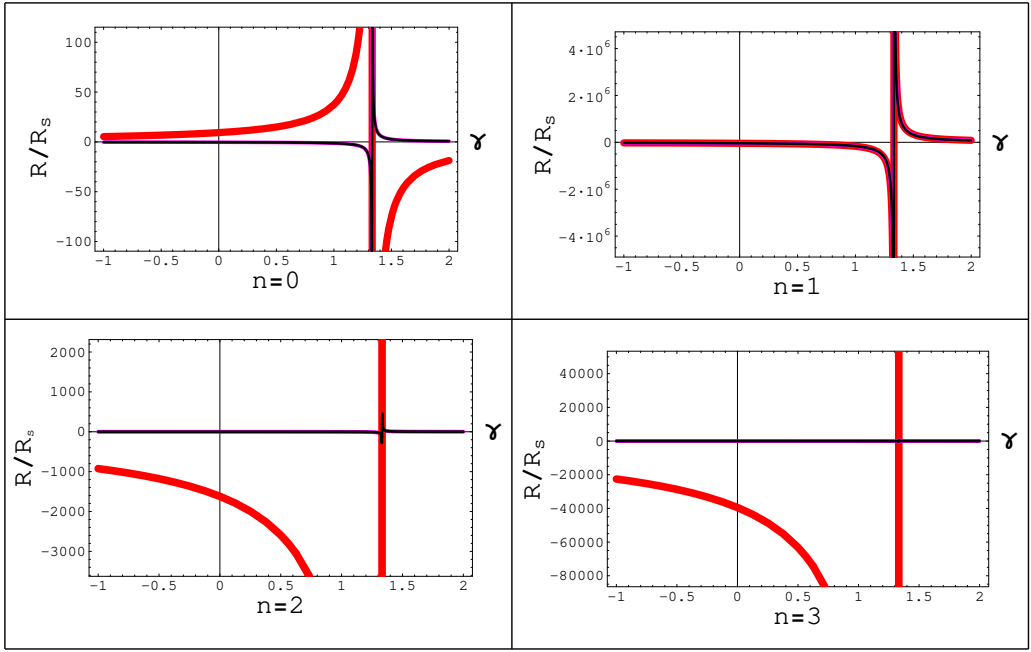,width=1\linewidth}
\caption{The values of $R/R_{s}=K/(\gamma-4/3)$ is plotted against
$(-1\leq\gamma\leq 2)$ for n=0,~1,~2,~3 in MBD, BD and GR
frameworks. Here, we have fixed $\omega_{BD}=6,~m_{0}=200
m_{AU}=200\times 10^{-27}$ and $\beta=1$. The red map shows MBD
ranges, BD theory results are mapped in black while purple colour
represents GR limits.}
\end{figure}

\section{Conclusion}

According to observational and experimental surveys, stellar
configurations are running far away from each other with an
accelerating rate causing accelerating expansion in the universe. It
is believed that this is due to the presence of dark energy
(mysterious energy) in the universe. Thus we cannot ignore the role
of dark energy in the evolution of stellar structure. Among
different dark energy candidates, BD gravity is considered as the
first prototype and the most fascinated alternative theory of
gravity. In order to be consistent with observational data, the BD
gravity is generalized (refines to) to MBD gravity (dilaton gravity)
in which the scalar field becomes massive and dilatons are
self-interacting due to the presence of potential of scalar field.

In this paper, we have discussed stability of spherical gaseous
masses for radial oscillations in the presence of dark energy by
incorporating MBD gravity. For this purpose, we have calculated
complete pN corrected hydrodynamics of MBD gravity in terms of
potential and super-potential functions. It is found that the
obtained solutions use some generalized potential functions that are
not involved in GR and BD gravity. This implies that stellar
configurations described by MBD gravity are more massive (have more
potential) than those of GR and BD theory. In order to discuss
radial oscillations of the system, we have perturbed the system by
Lagrangian radial perturbation and obtained linearized perturbed
dynamical equations. By applying variational principle on governing
perturbed equation of motion, we have formulated the criteria of
onset of dynamical instability for a special case $\gamma=constant$.
It is found that the results obtained for Newtonian approximation
are consistent with those described by GR and BD gravity but are
modified in pN correction.

In order to discuss realistic models in MBD theory, we have
evaluated radius of instability for different polytropic structures.
The resultant models depend upon the mass of scalar field and
provide a drastic change in the results of GR and BD gravity. The
system for $n=0$ is less stable than the systems described by GR and
BD. For $n=1$, the system is less stable than GR but more stable
than BD system. Polytropes for $n=1.5$ are less stable in MBD than
in GR. Structures for $n=2, 3$ are more stable in MBD gravity than
those described by GR and BD theory. For $n=0$, the system can be
stable for $\gamma<4/3$ which is inconsistent with GR. The case
$n=5$ gives unphysical result while for $n>5$, the stability range
is either $\gamma<4/3$ or $\gamma<4/3$ depending upon the  behavior
of $K$. We have also investigated the effects of scalar field mass
($m_{0}\gtrsim 200 m_{AU}=200\times 10^{-27}$) on the stability
criteria. It is found that it does not affect the instability ranges
defined on $\gamma$. However, the massive scalar field changes the
magnitude of radii of instability from BD and GR theories. It can be
noticed from the above discussions that the dynamics of massive
scalar field (MBD gravity) affects the hydrostatic timescales of
stellar structures. This implies that presence of dark energy not
only causes expansion in the universe but affects the evolution of
stellar evolutions.

From the above analysis, it can be noticed that the MBD gravity is
better option than BD gravity as it describes the most general
description of stellar evolutions which can be reduced to simple BD
(in the limits $(m_{0}<<\frac{1}{\tilde{r}})$ and
$\frac{V_{0}}{\phi_{0}}\rightarrow0$,) as well as GR case
($(m_{0}<<\frac{1}{\tilde{r}},~ \frac{V_{0}}{\phi_{0}}\rightarrow0$
and $\omega_{BD}\rightarrow\infty$). The analysis in MBD theory
deals with all types of situations such as massive scalar field,
massless scalar field and zero scalar field.

\section*{Appendix A}

The potential functions $U_{\alpha;i\alpha},~W_{i}(\textbf{x})$ and
$Z_{i(BD)}$ in Eq.(\ref{15}) are defined by
\begin{eqnarray}\nonumber
U_{\alpha;i\alpha}&=&G_{eff}\int_{v}\rho(\tilde{\textbf{x}})v_{\alpha}(\tilde{\textbf{x}})
\frac{(x_{i}-\tilde{x}_{i})(x_{i}-\tilde{x}_{i})d\tilde{x}}{\mid
\textbf{x}-\tilde{\textbf{{x}}}\mid^{3}},\\\nonumber
W_{i}(\textbf{x})&=&v_{\alpha}\frac{\partial}{\partial
x_{\alpha}}\left(U_{i}-U_{j;ij}\right)=
-G_{eff}\int_{v}\rho({\tilde{\textbf{x}}})v_{\alpha}({\textbf{x}})v_{\alpha}({\tilde{\textbf{x}}})
\frac{(x_{i}-\tilde{x}_{i})d\tilde{x}}{\mid
\textbf{x}-\tilde{\textbf{x}}\mid^{3}}\\\nonumber
&-&G_{eff}\int_{v}\rho(\tilde{x})
\left[v_{i}(\textbf{x})v_{\alpha}({\tilde{\textbf{x}}})+v_{i}({\tilde{\textbf{x}}})v_{\alpha}(\textbf{x})\right]
\frac{(x_{\alpha}-\tilde{x}_{\alpha})d\tilde{x}}{\mid
\textbf{x}-\tilde{\textbf{x}}\mid^{3}}\\\nonumber
&+&3G_{eff}\int_{v}\rho({\tilde{\textbf{x}}})\left[v_{\alpha}({\textbf{x}})
v_{\beta}(\tilde{\textbf{x}})(x_{\alpha}-\tilde{x}_{\alpha})(x_{\beta}-\tilde{x}_{\beta})\right]
\frac{x_{i}-\tilde{x}_{i}}{\mid\textbf{x}-\tilde{\textbf{x}}\mid^{5}}d\tilde{x}
,\\\nonumber Z_{i(BD)}&=&-2\frac{\partial}{\partial x_{i}}\left((1
+\frac{e^{-m_{0}r}}{3+2\omega_{BD}+e^{-m_{0}r}})U+\frac{\Lambda_{BD}
r^{2}}{3}\right)+2v^{2}\left(\frac{\partial U}{\partial
x_{i}}\right.\\\nonumber&+&\left.\left.\frac{\partial}{\partial
x_{i}}(\frac{e^{-m_{0}r}}{3+2\omega_{BD}
+e^{-m_{0}r}}U)+\frac{\partial(\gamma_{BD}U)}{\partial
x_{i}}+\Lambda_{BD}\frac{\partial r^{2}}{\partial
x_{i}}\right)\right.\\\nonumber &+&\left.p\left(2\frac{\partial
U}{\partial x_{i}}+2\frac{\partial}{\partial
x_{i}}(\frac{e^{-m_{0}r}}{3+2\omega_{BD}+e^{-m_{0}r}}U)
+2\frac{\partial(\gamma_{BD}U)}{\partial
x_{i}}\right.\right.\\\nonumber
&+&\left.\frac{\Lambda_{BD}}{2}\frac{\partial r^{2}}{\partial
x_{i}}\right)+c^{2}\left(U-\frac{\Lambda_{BD}
r^{2}}{6}\right)\Lambda_{BD}\frac{\partial r^{2}}{\partial x_{i}}.
\end{eqnarray}
The values of $Z_{ih(BD)}$ are given by
\begin{eqnarray}\nonumber
Z^{h}_{i(BD)}&=&-2\frac{\partial}{\partial x_{i}}\left((1
+\frac{e^{-m_{0}r}}{3+2\omega_{BD}+e^{-m_{0}r}})U
+\frac{\Lambda_{BD} r^{2}}{3}\right)\\\nonumber&+&\left.p
\left(2\frac{\partial U}{\partial x_{i}}+2\frac{\partial}{\partial
x_{i}}(\frac{e^{-m_{0}r}}{3+2\omega_{BD}+e^{-m_{0}r}}U)
+2\frac{\partial(\gamma_{BD}U)}{\partial
x_{i}}\right.\right.\\\nonumber
&+&\left.\frac{\Lambda_{BD}}{2}\frac{\partial r^{2}}{\partial
x_{i}}\right)+c^{2}\left(U-\frac{\Lambda_{BD}
r^{2}}{6}\right)\Lambda_{BD}\frac{\partial r^{2}}{\partial x_{i}}.
\end{eqnarray}
The values of $\Delta Z_{i(BD)}$ are
\begin{eqnarray}\nonumber
\Delta Z_{i(BD)}&=&-2 \frac{\partial}{\partial x_{i}}\left((1
+\frac{e^{-m_{0}r}}{3+2\omega_{BD}+e^{-m_{0}r}})\Delta
U\right)+\Delta p\left[2\frac{\partial}{\partial x_{i}}
U\right.\\\nonumber &+&\frac{2\partial}{\partial
x_{i}}(\frac{e^{-m_{0}r}}{3+2\omega_{BD}+e^{-m_{0}r}}
U)+\left.\frac{2\partial}{\partial
x_{i}}(\gamma_{BD}U)+\frac{\Lambda_{BD}}{2}\frac{\partial
r^{2}}{\partial x_{i}}\right]\\\nonumber &+&
p\left[\frac{2\partial}{\partial x_{i}}\Delta
U+2\frac{\partial}{\partial
x_{i}}\left(\frac{e^{-m_{0}r}}{3+2\omega_{BD}}\Delta
U\right)\right]+c^{2}\Lambda_{BD}\Delta U\frac{\partial
r^{2}}{\partial x_{i}}.
\end{eqnarray}

\end{document}